\documentclass[12pt,a4paper]{article}
\usepackage{a4wide}
\usepackage{graphicx}

\begin{document}

\begin{center}
{\bf Microelectronics Reliability {\bf 40 } (2000) 1781--1785.}
\end{center}

\vskip 2cm

\begin{center}
{\large {\bf MODELS FOR GENERATION }}$1/f${\large {\bf \ NOISE }}
\\[1.25\baselineskip] {\bf B.\thinspace Kaulakys$^{*,\dag }$ and
T.\thinspace Me\v skauskas$^{*,\ddag }$ }\\ {\small $^{*}$Institute of
Theoretical Physics and Astronomy, Vilnius University, A.\thinspace Go\v stauto 12, LT-2600
Vilnius, Lithuania }\\[0.25\baselineskip] {\small $^{\dag }$Faculty of
Physics, Vilnius University, Saul\.etekio al.\ 9, LT-2040 Vilnius, Lithuania
}\\[0.25\baselineskip] {\small $^{\ddag }$Faculty of Mathematics and
Informatics, Vilnius University, Naugarduko 24, LT-2006 Vilnius, Lithuania }
\\[0.25\baselineskip] {\small e-mail: {\sl kaulakys@itpa.lt}}
\\[1.25\baselineskip]
\end{center}

\begin{abstract}

Simple analytically solvable models are proposed exhibiting $1/f$ spectrum
in wide range of frequency. The signals of the models consist of pulses
(point process) which interevent times fluctuate about some average value,
obeying an autoregressive process with very small damping. The power
spectrum of the process can be expressed by the Hooge formula. The proposed
models reveal possible origin of $1/f$ noise, i.e., random increments of the
time intervals between pulses or interevent time of the process (Brownian
motion in the time axis).
\end{abstract}

\section{Introduction}

Widespread occurrence of signals exhibiting power spectral density with $1/f$
behavior suggests that a general mathematical explanation of such an effect
might exist. However, physical models of $1/f$ noise in some physical
systems are usually very specialized and they do not explain the
omnipresence of the processes with $1/f^\delta $ spectrum. Mathematical
algorithms and models for the generation of the processes with $1/f$ noise
can not, as a rule, be solved analytically and they do not reveal the origin
as well as the necessary and sufficient conditions for the appearance of $
1/f $ type fluctuations.

History of the progress in different areas of physics indicates to the
crucial influence of simple models on the advancement of the understanding
of the main points of the new phenomena. We note here only the decisive
influence of the Bohr model of hydrogen atom on the development of the
quantum theory, the role of the Lorenz model as well as the logistic and
standard (Chirikov) maps for understanding of the deterministic chaos and
the quantum kicked rotator for the revealing the quantum localization of
classical chaos.

Here we present simple models of $1/f$ noise which may essentially influence
on the understanding of the origin and main properties of the effect. Our
models are the result of a search for necessary and sufficient conditions
for the appearance of $1/f$ fluctuations in simple systems affected by the
random external perturbations, which where initiated in \cite{KV95C,KM97C}
and originated from the observation of the transition from chaotic to
nonchaotic behavior in the ensemble of randomly driven systems \cite
{KV95,KIM99}.

\section{Simple model}

The simplest version of our model corresponds to one particle moving along
some orbit. The period of this motion fluctuates (due to external random
perturbations of the system's parameters) about some average value $\bar
\tau $. So, a sequence of the transit times $\left\{ t_k\right\} $ when the
particle crosses some point of the orbit is described by the iterative
equations
$$
\left\{
\begin{array}{ll}
t_k= & t_{k-1}+\tau _k, \\
\tau _k= & \tau _{k-1}-\gamma \left( \tau _{k-1}-\bar \tau \right) +\sigma
\varepsilon _k.
\end{array}
\right. \eqno{(1)}
$$
Here $\gamma $ is the coefficient of the relaxation of the period to the
average value $\bar \tau $, $\left\{ \varepsilon _k\right\} $ denotes the
sequence of random variables with zero expectation and unit variance and $
\sigma $ is the standard deviation of the noise. Due to the contribution of
a large number of random variables to the transit times, model (1)
represents a long-memory random process.

Power spectral density $S\left( f\right) $ of the signal or current of the
model (1), $I\left( t\right) =a\sum_k\delta \left( t-t_k\right) $ (with $a$
being a contribution to the signal of one pulse or contribution to the
current of one particle when it crosses the section of observation), may be
calculated according to equation
$$
S\left( f\right) =\lim \limits_{T\rightarrow \infty }\left\langle \frac{2a^2}
T\left| \sum_{k=k_{\min }}^{k_{\max }}e^{-i2\pi ft_k}\right| ^2\right\rangle
\eqno{(2)}
$$
where $T$ is the whole observation time interval, $k_{\min }$ and $k_{\max }$
are minimal and maximal values of index $k$ in the interval of observation
and the brackets $\left\langle ...\right\rangle $ denote the averaging over
the realizations of the process. Eq. (2) may be rewritten in the form
$$
S\left( f\right) =\lim \limits_{T\rightarrow \infty }\left\langle \frac{2a^2}
T\sum_{k,q}e^{i2\pi f\Delta \left( k;q\right) }\right\rangle \eqno{(3)}
$$
where $\Delta \left( k;q\right) \equiv t_{k+q}-t_k$ is the difference of
transit times $t_{k+q}$ and $t_k$.

At $k\gg \gamma ^{-1}$ we have the stationary process \cite{KM98}: the
expectation $\left\langle \tau _k\right\rangle =\bar \tau $ and the variance
$\sigma _\tau ^2\equiv \left\langle \tau _k^2\right\rangle -\left\langle
\tau _k\right\rangle ^2=\sigma ^2/2\gamma $ of the recurrence time $\tau _k$
do not depend on the time while $\Delta \left( k;q\right) $ is a normally
distributed random variable with the expectation $\mu _\Delta \left(
q\right) \equiv \left\langle \Delta \left( \infty ;q\right) \right\rangle
=q\bar \tau $ and the variance $\sigma _\Delta ^2\left( q\right) \equiv
\left\langle \Delta \left( \infty ;q\right) ^2\right\rangle -\left\langle
\Delta \left( \infty ;q\right) \right\rangle ^2$ expressed as \cite{KM98}
$$
\sigma _\Delta ^2\left( q\right) =\left( \frac \sigma \gamma \right)
^2\left[ q-\frac{\left( 1-\left( 1-\gamma \right) ^q\right) }\gamma \right].
\eqno{(4)}
$$
For the normal distribution of $\Delta \left( k;q\right) =\Delta \left(
q\right) $ Eq. (3) yields
$$
S\left( f\right) =2\bar Ia\sum_q{\rm e}^{i2\pi f\mu _\Delta \left( q\right)
-2\pi ^2f^2\sigma _\Delta ^2\left( q\right) }\eqno{(5)}
$$
where $\bar I=\lim \limits_{T\rightarrow \infty }a\left( k_{\max }-k_{\min
}+1\right) /T=a/\bar \tau $ is the average current.

Substitution of the expansions of the variance (4) at $\left| q\right| \ll
\gamma ^{-1}$ in powers of $\gamma \left| q\right| $,
$$
\sigma _\Delta ^2\left( q\right) =\sigma _\tau ^2q^2,
$$
into Eq. (5) yields $1/f$-like power spectrum,

$$
S\left( f\right) =\bar I^2\frac{\alpha _H}f,\eqno{(6)}
$$
for sufficiently small parameters $\sigma $ and $\gamma $ in any desirably
wide range of frequencies, $f_1=\gamma /\pi \sigma _\tau <f<f_2=1/\pi \sigma
_\tau $. Here $\alpha _H$ is a dimensionless constant (the Hooge parameter),
$$
\alpha _H=\frac 2{\sqrt{\pi }}Ke^{-K^2},\quad K=\frac{\bar \tau }{\sqrt{2}
\sigma _\tau }.\eqno{(7)}
$$
We see that the power of $1/f$ noise except the squared average current $
\bar I^2$ depends strongly on the ratio of the average recurrence time $\bar
\tau $ to the standard deviation of the recurrence time $\sigma _\tau $.

Therefore, the process (1) containing only one relaxation time $\gamma ^{-1}$
\ can for sufficiently small damping $\gamma $\ produce an exact $1/f$-like
spectrum in wide range of frequency $\left( f_1,f_2\right) $, with $
f_2/f_1\simeq \gamma ^{-1}$.

The model is free from the unphysical divergency of the spectrum at $
f\rightarrow 0$. So, using for $f<f_1$ an expansion of expression (4) at $
\left| q\right| \gg \gamma ^{-1}$, $\sigma _\Delta ^2(q)=\left( \sigma
/\gamma \right) ^2\left( \left| q\right| -1/\gamma \right) $, we obtain from
Eq. (5) the Lorentzian power spectrum \cite{KM98}
$$
S\left( f\right) =\bar I^2\frac{4\tau _{rel}}{1+\tau _{rel}^2\omega ^2}.
\eqno{(8)}
$$
Here $\omega =2\pi f$ and $\tau _{rel}=D_t=\sigma ^2/2\bar \tau \gamma ^2$
is the ''diffusion'' coefficient of the time $t_k$. For $f\ll f_0=\bar \tau
\gamma ^2/\pi \sigma ^2=1/2\pi \tau _{rel}$ we have the white noise
$$
\begin{array}{c}
S(f)=\bar I^2\left( 2\sigma ^2/\bar \tau \gamma ^2\right) , \\
f\ll \min \left\{ f_1,f_0=\bar \tau \gamma ^2/\pi \sigma ^2\right\}.
\end{array}
$$
This is in agreement with the statement \cite{Schick74} that the power
spectrum of any pulse sequence is white at low enough frequencies.

Equations (6)--(8) describe quite well the power spectrum of the random
process (1) for perturbation by the Gaussian white noise and even for
perturbations by the non-Gaussian sequence of random impacts $\left\{
\varepsilon _k\right\} $ in Eq. (1) (see illustrative examples in Refs. \cite
{KM98} and \cite{KM98VC}).

\section{Generalizations and numerical analysis}

This simple exactly solvable model can easily be generalized in different
directions: for large number of particles moving in similar orbits with
coherent (identical for all particles) or independent (uncorrelated for
different particles) fluctuations of the periods, for non-Gaussian or
continuous perturbations of the systems' parameters and for spatially
extended systems. So, when an ensemble of $N_c$ particles moves on closed
orbits and the period of each particle fluctuates independently (due to the
perturbations by uncorrelated sequences of random variables $\{\varepsilon
_k^\upsilon \}$, different for each particle $\upsilon $) the power spectral
density of the collective current $I$ of all particles can be calculated by
the above method \cite{KM98} too and it is expressed as the Hooge formula
\cite{Hoo69,Hoo94}
$$
S\left( f\right) =\bar I^2\frac{\alpha _H}{N_cf},\quad \quad N_c=Vn_c.
\eqno{(6a)}
$$

This expression with the factor $1/N_c$ is in agreement with the Hooge \cite
{Hoo94,Hoo97} statements that summation of spectra is only allowed if the
processes contributing to the spectrum are isolated from each other and that
only isolated traps yield $1/f$ spectrum.

It should be noted that in many cases the intensity of signals or currents
can be expressed as a sequence of the pulse occurrence times $\left\{
t_k\right\} $, i.e., as $I\left( t\right) =a\sum_k\delta \left( t-t_k\right)
$. This expression represents exactly the flow of identical point objects:
cars, electrons, photons and so on. More generally, instead of the Dirac
delta function one should introduce time dependent pulse amplitudes $
A_k\left( t-t_k\right) $. The low frequency power spectral density, however,
depends weakly on the shapes of the pulses \cite{Schick74}, while
fluctuations of the pulses amplitudes result, as a rule, in white or
Lorentzian, but not $1/f$, noise. Model (1) in such cases represents
fluctuations of the averaged interevent time $\tau _k$ between the
subsequent occurrence times of the pulses.

The model may also be generalized for the nonlinear relaxation of the
interevent time $\tau _k$. In such a case Eq. (1) can be written in the form
$$
\left\{
\begin{array}{ll}
t_k= & t_{k-1}+\tau _k, \\
\tau _k= & \tau _{k-1}-\frac{dV(\tau _{k-1})}{d\tau _{k-1}}+\sigma
\varepsilon _k.
\end{array}
\right. \eqno{(9)}
$$
Here the function $V(\tau _k)$ represents the effective ''potential well''
for the Brownian motion of the interevent time $\tau _k$. The steady state
distribution density of the interevent time $\tau _k$ generated by Eq. (9)
is of the form
$$
\psi _\tau ^{{\rm st}}(\tau _k)=C\exp \left[ -\frac{2V(\tau _k)}{\sigma ^2}
\right] \eqno{(10)}
$$
where a constant $C$ may be obtained from the normalization. For the
power-law ''potential well''
$$
V(\tau _k)=\frac 12\gamma \left( \tau _k-\bar \tau \right) ^{2n}\eqno{(11)}
$$
with integer $n$ we have a generalization of Eqs. (1)
$$
\begin{array}{c}
t_k=t_{k-1}+\tau _k, \\
\tau _k=\tau _{k-1}-\gamma n\left( \tau _{k-1}-\bar \tau \right)
^{2n-1}+\sigma \varepsilon _k.
\end{array}
\eqno{(12)}
$$
The steady state distribution density of the interevent time $\tau _k$ in
such a case is
$$
\psi _\tau ^{{\rm st}}(\tau _k)=\frac{\left( \gamma /\sigma ^2\right)
^{1/2n}\exp [-\frac{\gamma \left( \tau _k-\bar \tau \right) ^{2n}}{\sigma ^2}
]}{2\Gamma \left( 1+1/2n\right) }.\eqno{(13)}
$$
For sufficiently large $n\gg 1$ Eqs. (12) and (13) represent Brownian motion
of the interevent time $\tau _k$ in almost rectangle ''potential well''
(square-well potential) restricting movement of $\tau _k$ mostly in the
interval $\left( \bar \tau -h,\bar \tau +h\right) $ with $h\simeq \left(
\sigma ^2/\gamma \right) ^{1/2n}$. At $h<\bar \tau $ such a restriction
prevents from the emergence of the negative interevent times $\tau _k$ and,
consequently, from the clustering of the particles or of the signal pulses.

We can evaluate the power spectrum of the processes (9) and (12) as well
\cite{KM98,K99C}. The power spectral density (3) may be rewritten in the
form
$$
S\left( f\right) =2\bar Ia\left\langle \sum_qe^{i2\pi f\tau _k\left(
q\right) q}\right\rangle \eqno{(14)}
$$
where the occurrence times $t_{k+q}$ and $t_k$ difference $\Delta \left(
k;q\right) $ is expressed as
$$
\Delta \left( k;q\right) \equiv t_{k+q}-t_k=\sum\limits_{l=k+1}^{k+q}\tau
_k=\tau _k\left( q\right) q,q\geq 0
$$
with $\tau _k\left( q\right) \equiv \left( t_{k+q}-t_k\right) /q$ being the
average interevent time in the time interval ($t_k,t_{k+q}$) and the
brackets in Eq. (14) denote averaging over time (index $k$) and over
realizations of the process. At $2\pi f\tau _k\left( q\right) \ll 1$ the
summation in (14) may be replaced by the integration
$$
S\left( f\right) =2\bar Ia\int\limits_{-\infty }^{+\infty }\left\langle
e^{i2\pi f\tau _kq}\right\rangle dq.
$$
Here the averaging over $k$ and over realizations of the process coincides
with the averaging over the distribution of the interevent times $\tau _k$,
i.e.,
$$
\left\langle e^{i2\pi fq\tau }\right\rangle =\int\limits_{-\infty }^{+\infty
}e^{i2\pi fq\tau }\psi _\tau \left( \tau \right) d\tau =\chi _\tau \left(
2\pi fq\right)
$$
where $\psi _\tau \left( \tau _k\right) $ is the distribution density of the
interevent times $\tau _k$ and $\chi _\tau \left( \vartheta \right) $ is the
characteristic function of this distribution.

Taking into account the property of the characteristic function
$$
\int\limits_{-\infty }^{+\infty }\chi _\tau \left( \vartheta \right)
d\vartheta =2\pi \psi _\tau \left( 0\right)
$$
we obtain the final expression for the power spectrum

$$
S\left( f\right) =2\bar I^2\bar \tau \psi _\tau \left( 0\right) /f.
\eqno{(15)}
$$
For Gaussian distribution of the interevent times $\tau _k$
$$
\psi _\tau \left( 0\right) =\exp \left( -\bar \tau ^2/2\sigma _\tau
^2\right) /\sqrt{2\pi }\sigma _\tau
$$
and according to Eq. (15) we recover expressions (6) and (7) obtained from
the analysis of the dynamical process.

Therefore, according to Eqs. (6), (7) and (15) the dimensionless parameter $
\alpha _H=2\bar \tau \psi _\tau \left( 0\right) $ is proportional to the
distribution density $\psi _\tau \left( 0\right) $ of the interevent time $
\tau _k$ in the point $\tau _k=0$, i.e., to the probability of the
clustering of the signal pulses. The pulse clustering results in the large
variance of the signal -- the condition necessary for appearance of
stationary $1/f$ noise in the wide range of frequency.

It should be noted, however, that Eq. (15) represent an idealized $1/f$
noise. The real systems have finite relaxation time and, therefore,
expression of the noise intensity in the form (14)--(15) is valid only for $
f>\left( 2\pi \tau _{rel}\right) ^{-1}$ with $\tau _{rel}$ being the
relaxation time of the interevent time's $\tau _k$ fluctuations. On the
other hand, due to the deviations from the approximation $t_{k+q}-t_k=\tau
_kq$ at large $q$, for sufficiently low frequency we can obtain the finite
intensity of $1/f^\delta $ ($\delta \simeq 1$) noise even in the case $\psi
\left( 0\right) =0$ but for the signals with fluctuations resulting in the
dense concentrations of the pulse occurrence times $t_k$.

We can generate, of course, the stationary time series of the occurrence
times $t_k$ also for other restrictions for the interevent time $\tau _k$,
e.g., with the reflecting boundary conditions at some values $\tau _k=\tau
_{\min }$ and $\tau _k=\tau _{\max }$. The process like (1) with the
reflecting condition for $\tau _k$ at $\tau _{\min }=\tau _{{\rm s}}$ may
also be generated by the recurrent equations
$$
\left\{
\begin{array}{ll}
t_k= & t_{k-1}+\tau _{
{\rm s}}+\tau _k \\ \tau _k= & \left| \tau _{k-1}-\gamma (\tau _{k-1}-\bar
\tau )+\sigma \varepsilon _k\right| .
\end{array}
\right. \eqno{(16)}
$$

Numerical analysis of the models like (1), (9), (12) and (16) as well as
with other restrictions for the interevent time $\tau _k$ shows that power
spectrum of the current is $1/f$-like in large interval of frequency only
when the distribution density of the interevent times $\tau _k$ in the point
$\tau _k\simeq 0$ is nonzero, i.e., $\psi _\tau (\tau _k\simeq 0)\neq 0$.
For models with $\psi _\tau (0)=0$ or $\psi _\tau (0)$ very close to zero we
observe in numerical simulations the power spectrum $S\left( f\right)
\propto 1/f^{3/2}$ (see Ref. \cite{KM98VC} and Fig. 1).

\section{Conclusions}

Simple analytically solvable models of $1/f$ noise are proposed. The models
reveal main features, parameter dependences and possible origin of $1/f$
noise, i.e., random increments of the time intervals between the pulses or
interevent times of the elementary events of the process. The conclusion
that $1/f^\delta $ noise with $\delta \simeq 1$ may result from the
clustering of the signal pulses, particles or elementary events can be drawn
from the analysis of the simple, exactly solvable models. The mechanism of
the clustering depends on the system.

\section*{Acknowledgments}

The authors acknowledge stimulating discussions with Prof. A. Bastys, Prof.
R. Katilius and Prof. A. Matulionis.

\nocite{*} \bibliographystyle{IEEE}

\newpage

\begin{figure}[tb]
   \begin{center}
      \includegraphics[width=0.7\hsize]{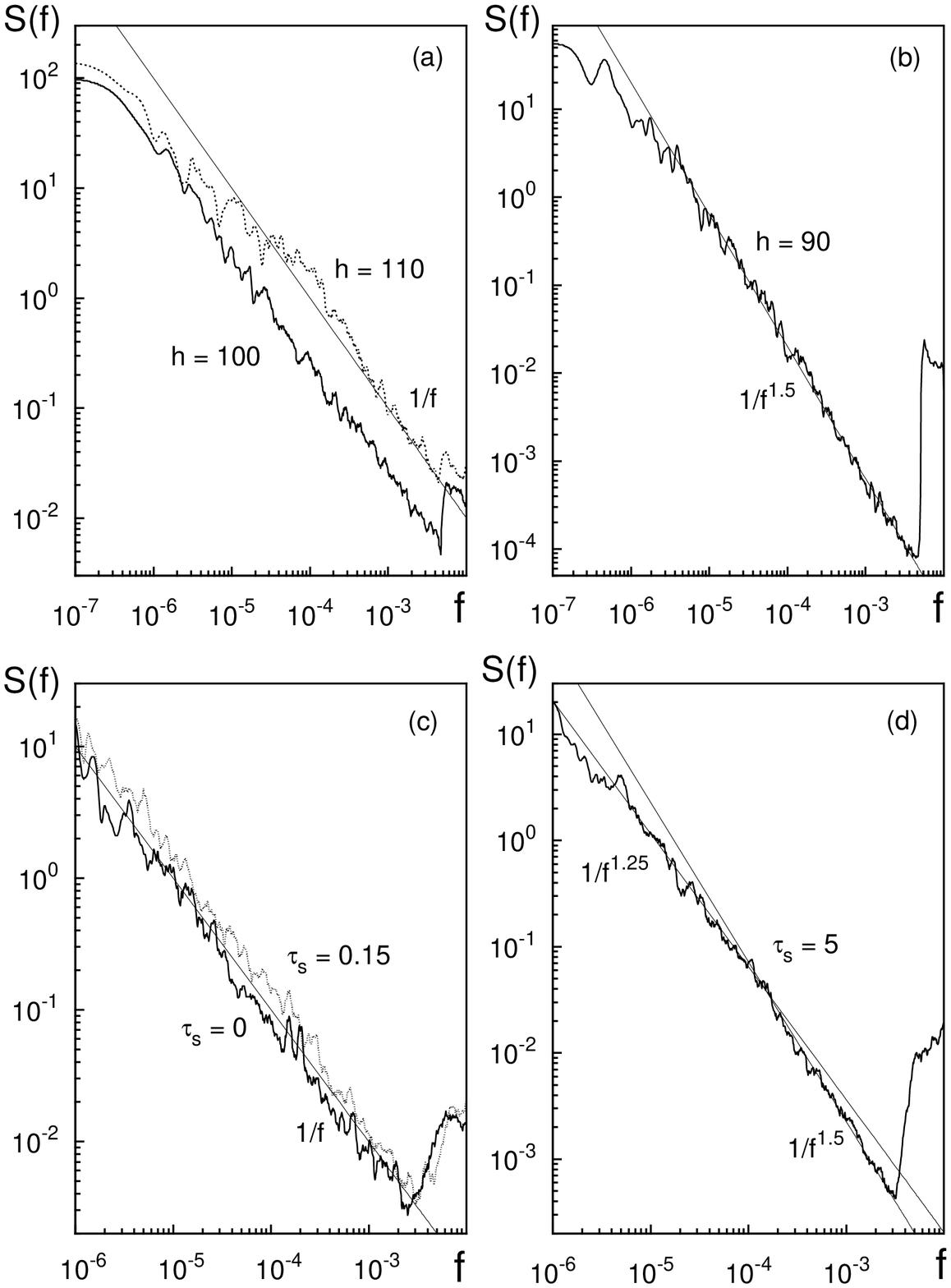}
   \end{center}
   {\small Fig.\,1.
    Power spectrum vs frequency of the current generated by Eqs.
(1)--(3) and (16) with parameters  $\bar \tau =100$, $\sigma =1$ and
with the Gaussian distribution of the random increments $\left\{
\varepsilon _k\right\} $. The sinuous curves represent the results of
numerical simulations:
(a) and (b) according to Eq. (1) with $\gamma =0$ and with
reflecting boundary conditions for $\tau _k$ at $\bar \tau - h$ and
at $\bar \tau + h$ for different values of $h$;
(c) and (d) according to Eq. (16) with $\gamma =0.0001$ and different
values of $\tau _s$.}
\end{figure}

\end{document}